\documentclass{article}
\usepackage[pdftex]{graphicx}
\graphicspath{ {figures/} }
\usepackage{array}
\usepackage{units}
\usepackage{fancyhdr}
\newlength\tindent
\usepackage{verbatim}
\usepackage[english]{babel}
\usepackage{cite}
\usepackage{float}
\usepackage{amssymb,amsmath,amsfonts}
\usepackage[T1]{fontenc}
\usepackage{hyperref}
\usepackage{enumerate}

\hypersetup{
  colorlinks   = true, 
  urlcolor     = blue, 
  linkcolor    = blue, 
  citecolor   = red 
}
\usepackage{ae,aecompl}
\usepackage{xspace}
\usepackage{multirow}
\usepackage{bigstrut}
\usepackage{soul}
\usepackage{caption}
\usepackage{enumerate}
\usepackage{mathrsfs}
\usepackage{titlesec}
\usepackage[export]{adjustbox}
\usepackage{subcaption} 
\usepackage{amsmath}
\usepackage[usenames,dvipsnames,svgnames,table]{xcolor}
\usepackage{blindtext}
\usepackage{scrextend}
\usepackage{graphicx}
\usepackage{mathtools,slashed}
\usepackage{subcaption}  
\usepackage{enumerate} 
\usepackage{amsmath,amsfonts,amssymb}  
\usepackage[toc,page]{appendix}

\usepackage[T1]{fontenc}
\usepackage{babel}
\usepackage{array}
\usepackage{diagbox,ragged2e}
\usepackage{fourier}

\newcommand{\RN}[1]{
  \textup{\uppercase\expandafter{\romannumeral#1}}
}

\usepackage{mathtools,slashed}

 \usepackage{enumitem}

\usepackage[nottoc,notlot,notlof]{tocbibind}
\usepackage{blindtext}
\title{Maximum Average Entropy-Based Quantization of Local Observations for Decentralized Detection
\thanks{ A preliminary version of this paper was presented in the 27th Signal Processing and Communications Applications Conference (SIU), Sivas, Turkey,  24-26 April 2019.
}}
\author{Muath A. Wahdan and Mustafa A. Alt{\i}nkaya
\and Department of Electrical and Electronics Engineering \\
\.{I}zmir Institute of Technology \\
\.{I}zmir, Turkey\\
muathwahdan@iyte.edu.tr and mustafaaltinkaya@iyte.edu.tr }
\begin{document}
\maketitle
\section{Abstract}
In a wireless sensor network the sensor outputs are required to be quantized because of energy and bandwidth requirements. We propose such a distributed detection scheme for a point source which is based on Neyman-Pearson criterion where sensor outputs are quantized by maximizing the average output entropy of the sensors under both hypotheses. The quantized local outputs are transmitted to a fusion center (FC) where they are equally combined to make  a global decision. The performance of the proposed maximum average entropy (MAE) method in quantizing sensor outputs was tested for binary, three-level, four-level and six-level quantization. 
The effects of the channel from the sensors to the FC is also addressed for the MAE by using both  orthogonal fading channel model and a simplified channel model in addition to the direct data transmissions. A Comparison between the proposed method MAE and J-divergence method  has been performed for the direct data transmission. The simulation studies show that  MAE outperform the J-divergence method.
\section{Introduction}
Wireless Sensor Networks (WSNs) have come into the spotlight recently due to a major development in the Micro-Electro-Mechanical Systems (MEMS) \cite{MEMS}. The recent development of WSNs have made this field a research focus of intensive researches. The researchers are widely using it in monitoring and characterizing large physical environments and for tracing various environmental or physical conditions such as temperature, pressure, wind, and humidity. Apart from these, WSNs have vast fields to be applied in, such as harmful environmental exploration, wildlife monitoring, relief from natural catastrophe and military target tracking and surveillance \cite{Application1, Application2, Application3}. 
 
Typically a WSN uses a huge number of comparatively inexpensive and low-energy sensors to collect observations and pre-process the observations. These sensors are always deployed in the environment. Usually, each of these sensor nodes has the capability to communicate with other sensor nodes or the base station (fusion center) through a wireless channel. Typically,  the fusion center (FC) will be responsible to collect data from sensor nodes and perform a global decision.
 
In our work, we will concentrate on the decentralized detection problem of a WSN.
 Owing to strict energy and bandwidth restrictions, observations of the sensors are frequently needed to be quantized, before transmitting them to a fusion center (FC) where a global decision is made \cite{Ciuonzo, Al-Jarrah}.
 Distributed detection systems \cite{Ekchian} show the advantages of higher survivability and reliability than their centralized counterparts.
In \cite{1}, the authors consider a detection problem consisting of two sensors and one FC with a fixed fusion rule to show that the optimum local decision rule is the likelihood ratio test under the Bayesian criterion. Then, in \cite{2} and \cite{3}, it was shown that the optimum fusion rule at the FC is also a likelihood ratio test both under the Neyman-Pearson and the Bayesian criteria. 
Optimum quantization levels in the sense of information theoretic criteria for distributed detection systems were presented in \cite{quantization1,quantization2}. In \cite{quantization1}, J-divergence has been used to optimize the distributed detection of a serial system with two sensors for the Bayesian detection criterion. In \cite{quantization2}, optimum quantization levels have been investigated for a binary system under the assumption that the likelihood functions under both hypotheses and the binary decision threshold are given and fixed. Then by further partitioning of decision regions two bit decisions instead of 1 bit decisions are sent to the FC. In that work, it was assumed that all local sensors are identical NP detectors observing the same signal-to-noise ratio (SNR).
In \cite{quantization3}, the optimal quantization intervals based on deflection criterion (DC) and 
Chernoff information (CI) are defined for distributed detection systems consisting of one FC and multiple sensors by using Bayesian detection criterion for known SNR.

In this paper, we propose an entropy based method for determining the quantization intervals at
distributed sensors in order to optimize the global binary decision at the FC about the
existence of a point source under the Neyman-Pearson criterion where sensors observe different signal levels which they do not know and extend the preliminary version of this paper \cite{SIU} in the following aspects. We compare this method to popular J-divergence \cite{quantization1} based method, demonstrate the positively proportional relation of the proposed method with J-divergence, included increased quantization levels resulting in similar performance to non-quantized signalling. Instead of the binary symmetric channel as a simplified model for the channel from the sensors to the FC, we use regular Rayleigh fading channel model for the wireless channel. Additionally, modulation in this transmission is also accounted for. We consider scenarios with non-equivalent importance of hypotheses, that is why NP criterion is considered to be more suitable compared to probability of error criterion in this work.

This paper is organized as follows. First, we formulate the parallel distributed detection problem that includes the Rayleigh fading channel for various fusion rules in Section \ref{System model}.
 Section \ref{QUANTIZER DESIGN} describes the effect of quantizing the local observations on the system performance. Based on these results the application of the proposed method for binary, three-level,  four-level and six-level quantization of sensor observations for detecting a point source using a wireless sensor network (WSN) also considered. The experimental study and simulation results are given in Section \ref{Experimental Study}. Conclusions are drawn in Section \ref{section:Conclusion}.\\
Notations: Boldface lower and upper case denote vectors and matrices, respectively. The symbol $''\sim''$ denotes the distributed according to, whereas $\mathcal{N}({\mu},\,\sigma^2)$ denotes Gaussian pdf with mean ${\mu}$ and variance $\sigma^2$. $C\mathcal{N}(\boldmath{\mu},\,\textbf{C}_{\bar{\textbf{y}}})$ denotes complex Gaussian pdf with mean vector $\boldsymbol{\mu}$ and covariance matrix $\mathbf{C}$.



\section{System model}\label{System model}
A binary hypothesis testing problem has been considered in this work, where a group of $K$ sensors and one FC cooperate to detect the existence of a point source as shown in Figure \ref{binary hypothesis testing}. The hypothesis testing at each sensor node can be described as
\begin {equation}
\begin{split}
\label{binary hypothesis testing}
H_0 &: y_k= \epsilon_k,\\
\text {versus}\\
 H_1&: y_k = A_k + \epsilon_k,
\end{split}
\end{equation}
where $y_k \in {\rm I\!R}$ denotes the  observation at the $k$th sensor,
$\epsilon_k  \in {\rm I\!R}$ denotes additive white Gaussian noise (AWGN) with variance $\sigma^{2}$ and zero mean and $ {\rm I\!R}$  is the set of the real numbers. $A_k $ denotes the received signal amplitude which is equal to $\alpha_k A_{max}$. Each sensor in the range of the point source detect a signal attenuated with a factor of  $\alpha_k$ and make a local decision $u_k$. The local decision is transmitted through multiplicative channel $h_k$ to the FC where the final decision $u_0$ is made. ${\bar{y}_k}$ denotes the received signal at FC, for $k = 1,2,...,K$.
\begin{figure}[H]
\centering
\includegraphics[width=12.0cm]{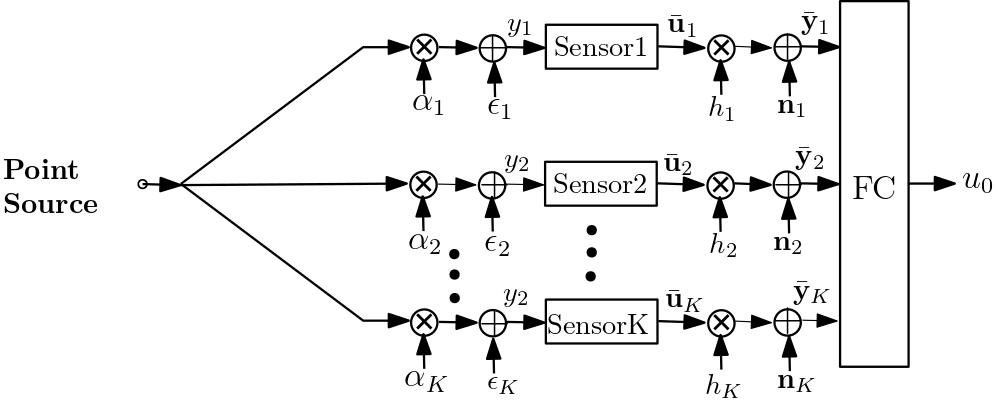}
\caption{Parallel distributed detection system. }
  \label{Parallelmodel_Fading1}
\end{figure}

Based on the dispersion pattern over the surveillance zone and the physical characteristics, the phenomenon to be detected can be modeled either as a field source or a point source. A field source is dispersed over the sensor field such as in temperature monitoring. On the other hand, the event is generated by a single point source such as in target detection and fire detection.

\subsection{Point source}
In our work, we consider a point event source emitting constant power uniformly in all directions. For such a source the signal amplitude received by a sensor will be inversely proportional to the distance from the source. Considering uniformly deployed sensors, only those sensors which are within a circle the radius of which is determined by the sensitivity of the sensors, will receive a signal.  

\begin{figure}[H]
\centering
\includegraphics[width=12.0cm]{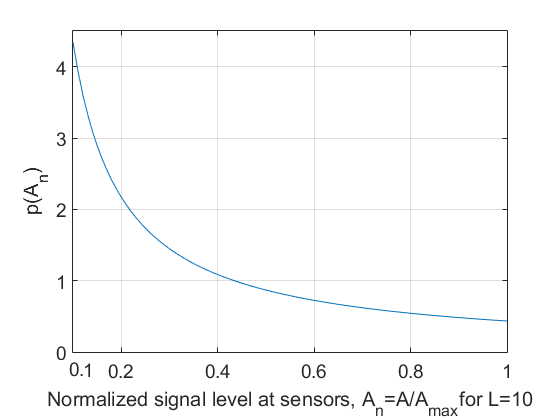}
\caption{The probability density function $p(A_n)$.}
  \label{fig:1}
\end{figure}

Let $A_\text{max}$ denote the signal amplitude on a circle with radius $r_{\text{min}}$ centered by the event location as shown in Figure \ref{circle_new2}.  We assume that $A_\text{max}$ corresponds to the maximum detectable signal level or the saturation level of the sensors and $A_\text{min}$ denotes the minimum value of the detectable signal observed at a distance of  $r_{\text{max}}$ from the event location. This yields a different and unknown amplitude value for each individual sensor. Assuming there are no sensors in the small circle, the pdf of the normalized signal amplitude, $A_n=A/A_\text{max}$, at a sensor will have the form shown  in Figure \ref{fig:1} and will be given as:

\begin{figure}
\centering
\includegraphics[width=10.0cm]{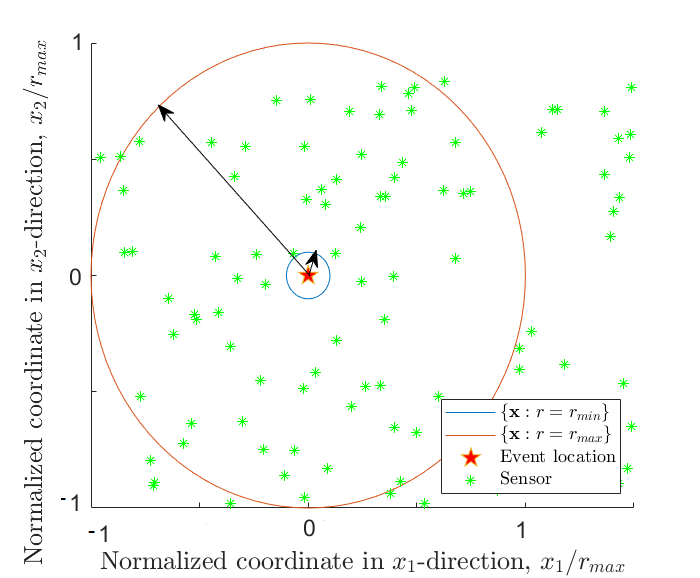}
\caption{Positions of the event location and uniformly distributed sensors transmitting to the FC. }
  \label{circle_new2}
\end{figure}

\begin{equation}
\label{pdf_of_A}
p{(A_n)}= \frac{1}{A_n\log(L)}
\end{equation}
where $L=A_\text{max}/A_\text{min}$ and $\log(\cdot)$ is the natural logarithm. We define the SNR  as the ratio between the maximum signal power, $A_{\text{max}}^2$, and the noise power, $\sigma^{2}$. Let us assume that $K$ of the sensors uniformly deployed in the area will be in the ring described by the radius $r_\text{min}$ and $r_\text{max}$. Then, the signal amplitudes at these sensors will be independent and come from the pdf given in (\ref{pdf_of_A}) in the case of an event. Assuming that the sensor observations are available distortion-free at the FC, i.e. without transmission over a wireless channel, the optimal Bayesian NP detector can be written as:\\
\begin{equation}
\label{eq:2.23}
\mathbf\Lambda(\textbf{y})=\frac{\displaystyle \prod_{k=1}^K\displaystyle \int_{A_\text{max}/L}^{A_\text{max}} p({y_k}|H_1; A_k) p(A_k) dA_k}{p(\textbf{y}|H_0) } \underset{H_0}{\stackrel{H_1}\gtrless}\eta.
\end{equation}
Since each $A_k$ comes from the independent and identical pdf given in (\ref{pdf_of_A}), we eliminate the index, $k$, and express the likelihood ratio as
\begin{equation}
\label{eq:2.24}
\mathbf\Lambda(\textbf{y})= \frac{\displaystyle \prod_{k=1}^K\displaystyle \int_{A_\text{max}/L}^{A_\text{max}} \frac{1}{\sqrt{2 \pi \sigma^{2}}} \exp\left( \frac{- (y_k - A)^{2}}{2 \sigma^2}\right) \frac{1}{A\log(10)} dA}{ \left(\frac{1}{\sqrt{2 \pi \sigma^{2}}}\right)^{K} \exp\left( \frac{- \sum_{k=1}^K(y_k )^{2}}{2 \sigma^2}\right)} \underset{H_0}{\stackrel{H_1}\gtrless}\eta,
\end{equation}%
where $\mathbf{y}=[y_1,y_2,...,y_K]$ denotes vector of observations from $K$ sensors.
 
\subsection{Fusion System: Channel Between Sensors and FC}\label{channel types}
In this section we will investigate the general case by using the frequency-shift keying (FSK) modulation scheme where digital data is transmitted by transmitting a carrier wave of different frequency for each different data item. $M$-FSK that is FSK with M different symbols is a suitable modulation scheme for low power low data rate data transmission as preferred by majority of the sensor device equipment. In this paper, $M$-FSK modulation scheme with non-coherent demodulation has been investigated for Rayleigh fading channels and AWGN. Training symbol transmit power is zero and different sizes of modulation constellation are utilized as described in subsection \ref{Fading channel}. Moreover, in order to concentrate on the fusion of sensor data with non-identical signal levels, we consider the case of no channel i.e. when error-free sensor outputs are available at the fusion center which we called as direct data transmission (DDT). Once data from the sensors are at the FC, an equal gain fusion rule is applied for every different types of sensor transmissions to FC since the relative reliability of sensor outputs are not evaluated.

\subsubsection{Fading channel}\label{Fading channel}

In this subsection, the problem of fusing the data transmitted over a fading channel has been considered, as shown in Figure \ref{Parallelmodel_Fading1}. The FC has only information on the channel statistics. Non-coherent $M$-FSK modulation is employed for transmitting data to the FC.
Let $\mathbf{u}_k$  denote the $M$-FSK modulated symbol at sensor $k$, where $\mathbf{u}_k \in\{\mathbf{e}_m, m=1, ..., M\}$ and $\mathbf{e}_m$ is an $M \times 1$ column vector, all elements of which except the $m$th one are zero. We refer to the transmit power of the data symbol as $P_d$. By assuming that  symbols are transmitted over orthogonal channels between the detectors and the FC, the output of the channel which is corresponding to detector $k$ at the FC is:

\begin{equation}
\label{MFSK}
\begin{split}
\bar{\mathbf{y}}_k&=\sqrt{P_k} h_k \mathbf{u}_k+\mathbf{n}_k\\
&= h_k \bar{\mathbf{u}}_k+\mathbf{n}_k
\end{split}
\end{equation}
where $P_k=P_d(\Delta_k)^{-\epsilon}\lambda^2/(4\pi)$ represents the received power \cite{OptimumFC}, $\lambda$ is the wavelength, $\epsilon$ is the path loss exponent, and $\Delta_k$ is the distance between detector $k$ and the FC. The channel noise is denoted as $\mathbf{n}_k$ which is a zero mean complex Gaussian vector $\mathbf{n}_k \sim  C\mathcal{N}(0,\,\sigma_n^{2}I)$, where $I$ is an $M \times M$ identity matrix. The complex channel coefficient $h_k$ in (\ref{MFSK}) is modeled as  $h_k \sim  C\mathcal{N}(0,1)$ which can be also represented as $h_k =\alpha_k e^{j\phi_k}$, where $\alpha_k$ represents the amplitude with Rayleigh distribution and $\phi_k$ represents the phase with uniform distribution. We adopt NP criterion to find the optimal and a sub-optimal fusion rule at the FC in order to obtain a global decision $u_0 \in \{H_0, H_1\}$ as follows.
 
\begin{figure}[H]
\centering
\includegraphics[width=13.0cm]{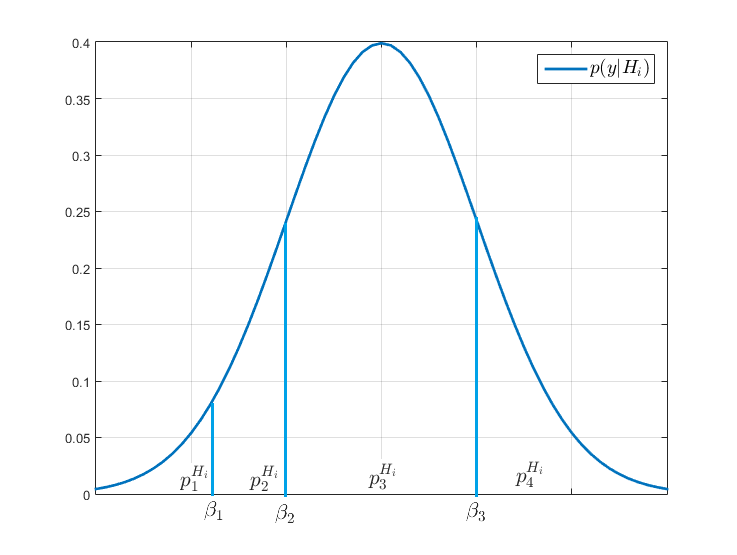}
\caption{A partitioning of the pdf for the observations at each sensor for 4-level quantization.}
  \label{Partitioning of the pdf for the observations at each sensor}
\end{figure}

\begin{enumerate}[label=(\roman*)]
  \item The optimal fusion rule for the i.i.d. vectors, $\bar{\textbf{y}}_k, k=1,2,...,K$, is defined as follows:

\begin{equation}
\label{fusion rule v1}
\log \mathbf\Lambda(\textbf{Y})=\log \frac{p(\textbf{Y}|H_1)}{p(\textbf{Y}|H_0)}= \log \prod_{k=1}^K \frac{p(\bar{\textbf{y}}_k|H_1)}{p(\bar{\textbf{y}}_k|H_0)}\underset{H_0}{\stackrel{H_1}\gtrless}\eta,
\end{equation}%
where $\mathbf{Y}$ is the matrix composed by row-wise stacking column vectors ${\mathbf{y}_k, k=1, ... , K}$

\begin{equation}
\label{fusion rule v2}
\log \mathbf\Lambda(\textbf{Y})= \sum_{k=1}^{K} \log  {\frac{p(\bar{\textbf{y}}_k|H_1)}{p(\bar{\textbf{y}}_k|H_0)}}\underset{H_0}{\stackrel{H_1}\gtrless}\eta.%
\end{equation}
Expanding $p(\bar{\textbf{y}}_k|H_0)$ and $p(\bar{\textbf{y}}_k|H_0)$ in (\ref{fusion rule v2}) over the $M$-level sensor decisions we obtain
\begin{equation}
\label{fusion rule v3}
\log \mathbf\Lambda(\textbf{Y})= \sum_{k=1}^{K} \log  \left(\frac{\sum_{m=1}^{M} p(\bar{\textbf{y}}_k|\textbf{u}_k(m)) p(\textbf{u}_k(m)|H_1)}{\sum_{m=1}^{M} p(\bar{\textbf{y}}_k|\textbf{u}_k(m)) p(\textbf{u}_k(m)|H_0)}\right)\underset{H_0}{\stackrel{H_1}\gtrless}\eta,
\end{equation}
 where $K$ represents the number of sensors and $M$ represents the number of quantization levels at each local sensor. The conditional density $p(\bar{\textbf{y}}_k|\textbf{u}_k(m))$ in (\ref{fusion rule v3}) is a complex multi-variate Gaussian density, $\bar{\textbf{y}}_k \sim  C\mathcal{N}(0,\,\textbf{C}_{\bar{\textbf{y}}})$, $\textbf{C}_{\bar{\textbf{y}}}$ represents the diagonal matrix with entries $\textbf{C}_{\bar{\textbf{y}}}(j,j)=\sigma_n^{2}$ for $j\neq m$ and $\textbf{C}_{\bar{\textbf{y}}}(j,j)=P_k \sigma_h^{2}+ \sigma_n^{2}$ for $j=m$, where $j=1, ..., M$. We can prove that  $p(\bar{\textbf{y}}_k|\textbf{u}_k(m))$ equals to
 
 \begin{equation}
\label{liklihood}
\frac{1}{\sqrt{\pi^M \det{|\mathbf{C}_{\bar{\mathbf{y}}_m}|}}} \exp{-(\bar{\textbf{y}}_k-\boldsymbol{\mu})^H \textbf{C}_{\bar{\mathbf{y}}_m}^{-1}(\bar{\mathbf{y}}_k-\boldsymbol{\mu}) }.
\end{equation}%
The values of $ p(\textbf{u}_k(m)|H_1)$ represents the probability masses under hypothesis $H_1$, which are estimated as:
   \begin{equation}
\label{prob1}
\overline{p^{H_1}_{m}}=\int_{A_{\text{max}/L}}^{A_{\text{max}}}{p_m^{H_1}(A_n) p(A_n)dA_n},
\end{equation}
where ${p_m^{H_1}(A_n)}$ represents the probability mass under $H_1$ as shown in Figure \ref{Partitioning of the pdf for the observations at each sensor} for an observed signal level $A_n$ which is the mean of the Gaussian signal.\\
The value of $ p(\textbf{u}_k(m)|H_0)$ represents the probability masses under hypothesis $H_0$:
  
  \begin{equation}
\label{prob2}
p(u_k(m)|H_0)=p^{H_0}_{m}.
\end{equation}
Figure \ref{Partitioning of the pdf for the observations at each sensor} shows the probability masses for $M=4$ under hypothesis $H_i$, $i=1,2$.

  \item A sub-optimal fusion rule can be derived as follows:
In (\ref{fusion rule v3}) we see both the effects of fading channel and the local detection outputs in order to achieve the optimal performance. A direct alternate could be used as a sub-optimal fusion rule by separating this into two-steps. First, $\bar{\mathbf{y}}_k$ is used to infer about the local detector by applying the maximum likelihood (ML) estimate  as an intermediate decision,$\hat{u}_k$, and then, the optimum fusion rule based on $\hat{u}_k$ is applied:
  
    \begin{equation}
\label{subOpt}
\hat{u}_k={\arg \max_{m}} \quad {\boldsymbol{\theta}_m},
\end{equation}%
  where ${\boldsymbol{\theta}_m}$ is given as
      \begin{equation}
\label{theta}
{\boldsymbol{\theta}_m}= p(\bar{\textbf{y}}_k|\textbf{u}_k(m)).
\end{equation}
   We can re-write (\ref{liklihood}) as in \cite{OptimumFC}
   
   \begin{equation}
\label{liklihood2}
p(\bar{\textbf{y}}_k|\textbf{u}_k(m))=\frac{1}{\sqrt{\pi^M \det{|\textbf{C}_{\bar{\mathbf{y}}_m}|}}} \exp\left({\frac{P_k \sigma_h^{2}|\bar{y}_k^m|^2}{\sigma_n^{2}(\sigma_h^{2}+ \sigma_n^{2})}}\right) \prod_{j=1}^M \exp\left({\frac{|\bar{y}_k^j|^2}{\sigma_n^{2}}}\right).
\end{equation}
By substituting (\ref{liklihood2}) in (\ref{subOpt}) after eliminating the terms which are irrelevant to $m$, we can re-write (\ref{subOpt}) as
      \begin{equation}
\label{subOpt1}
\hat{u}_k={\arg \max_{m}} \quad \exp\left({\frac{P_k \sigma_h^{2}|\bar{y}_k^m|^2}{\sigma_n^{2}(\sigma_h^{2}+ \sigma_n^{2})}}\right),
\end{equation}
 where $m=1, ..., M$. Note that $|\bar{y}_k^m|^2$ in (\ref{subOpt1}) denotes  the squared envelopes of $M$ cross-correlators corresponding to non-coherent FSK detection. 
\\
  The final decision rule is given as
   \begin{equation}
\label{finalDecsion}
{u_0}=\sum_{k=1}^K\hat{u}_k \underset{H_0}{\stackrel{H_1}\gtrless}\eta.
\end{equation}
\end{enumerate}

\clearpage

\section{QUANTIZER DESIGN}\label{QUANTIZER DESIGN}

It is aimed to make a global decision at the FC under the NP criterion. Let us assume that each
sensor will only make a single observation and will transmit this observation to the FC. Then, sensors will make i.i.d. observations under $H_0$ and none of the sensors can estimate the signal level under $H_1$. Consequently, there is no information at the sensors in order to use different quantization thresholds under $H_1$. So, it is reasonable to use identical quantization thresholds at each sensor irrespective of their distance to the event location since it cannot be estimated. Definitely, the choice of the quantization thresholds affects the performance which makes it desirable to choose the quantization thresholds which maximizes the system performance. This paper proposes maximum average entropy (MAE) method, that is, determining the quantization thresholds at the sensors in the way to maximize the average entropy of the discrete information collected at the FC under both hypotheses without considering the effects of the succeeding wireless channel. The optimum detector at the FC is based on likelihood ratios as given in $(\ref{eq:2.24})$. Equivalently, one can use log-likelihood (logarithm of likelihood) ratios and the log-likelihood ratio for the $k$th sensor and unknown signal amplitude can be calculated by using the expected value of the signal amplitude, $\bar{A}$, as follows:
 \begin{equation}
 \label{liklihoodEstiated}
 \log(\Lambda)=-\frac{\bar{A}^2}{2\sigma^{2}}+\frac{\bar{A}}{\sigma^{2}}y_k.
 \end{equation}
The linear (or more appropriately affine) transformation of observations in $(\ref{liklihoodEstiated})$ to log-likelihood ratios is irrelevant in entropy based quantization because that kind of transformation only results in translation and scaling of the underlying pdfs and will preserve the resulting probability masses corresponding to a vector of thresholds (such as $\beta_1$, $\beta_2$ and $\beta_3$ in Figure $\ref{Partitioning of the pdf for the observations at each sensor}$.
Another information based criterion for determining the quantization thresholds is  the maximum
J-divergence (MJD) method which was used in the case of constant signal level at sensors formerly  \cite{quantization2}. We will first explain these criteria and then  the relation between them.

\subsection{Maximum Average Entropy Method}
\begin{figure}
\centering
\includegraphics[width=12cm]{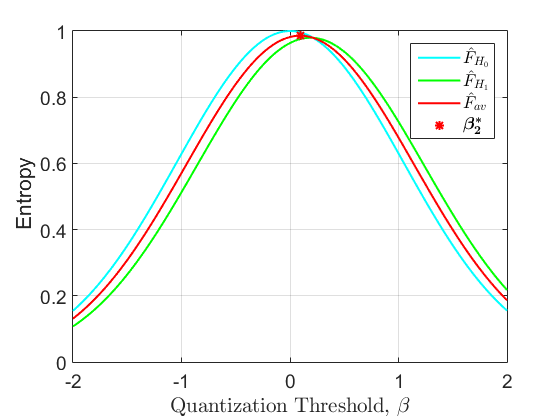}
\caption{ The entropy functions $\hat{F}_{H_0}$, $\hat{F}_{H_1}$ and 
  $\hat{F}_{\text{av}}$ for binary quantization.}
  \label{Binary entropy}
\end{figure}

 An intuitive idea to have an optimum performance at the FC is to maximize the entropy under both hypotheses which we call as Maximum Average Entropy (MAE) method \cite{SIU}.
 %
So, we propose to determine the quantization intervals at the sensors as resulting in MAE under both hypotheses. The entropy of a quantized sensor output can be calculated based on the partitioning of the observation pdf at each sensor as shown in Figure \ref{Partitioning of the pdf for the observations at each sensor}.
In this figure, the number of quantization intervals is $4$. For a general number of $M$ quantization intervals, there will be $M-1$ thresholds, $\{\beta_1, \beta_2, . . ., \beta_{M-1} \}$, and $M$ partitions with corresponding probability masses of likelihood ratios $[p_1^{H_i}, p_2^{H_i}, ..., p_M^{H_i} ]$ where $i=0,1$. Under $H_i$, the entropy of the observation can be estimated as 
 
\begin{equation}
\label{lrcm}
\hat{F}_{H_i}=\hat{E}\left( -\sum_{m=1}^M p_{m}^{H_i} \log_2(p_{m}^{H_i})\right)\text{bit}. 
\end{equation}%
The expectation, $\hat{E}(\cdot)$, is with respect to the distribution of the $K$ sensors and in the special case of the scenario described in Figure \ref{circle_new2}, this distribution is uniform  in the sensing range of the sensors defined by a circle within a radius of $r_\text{max}$ from the event location. $\mathbf {p}_M^{H_0}=[p_1^{H_0}, p_2^{H_0}, ..., p_M^{H_0} ]$ denotes the vector of these probability masses, i.e. the probabilities of the partitions.  In practice, an estimate of this expectation is obtained by averaging the information of the sensors over the distribution of the sensor locations and AWGN realizations which is called a histogram method \cite{entropy2}. Figure \ref{Binary entropy} shows the entropy function $\hat{F}_{H_0}$, $\hat{F}_{H_1}$ and $\hat{F}_{\text{av}}=\frac{1}{2}( \hat{F}_{H_0}+\hat{F}_{H_1})$ for binary quantization. For $M$-ary quantization, $\boldsymbol{\beta}^*_M=[\beta^*_1, \beta^*_2, . . ., \beta^*_{M-1}]$ denotes the vector of optimum quantization thresholds, in the sense of MAE which is found as 

\begin{equation}
\label{eq8}
\boldsymbol{{\beta}}^*_M={\arg \max_{\boldsymbol{\beta}_M}} \quad \hat{F}_{\text{av}}.
\end{equation}%

\begin{figure}
\centering
\includegraphics[width=12cm]{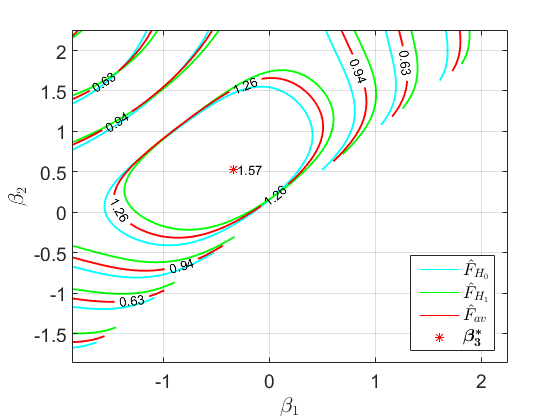}
\caption{The entropy functions $\hat{F}_{H_0}$, $\hat{F}_{H_1}$ and 
  $\hat{F}_{\text{av}}$ for three level quantization.}
  \label{Entropy_threeLevel}
\end{figure}

Optimal quantization thresholds for binary quantization corresponds to the maximum of $\hat{F}_{\text{av}}$ as shown  in Figure  \ref{Binary entropy} which is $\beta^*_2=0.093$.  In a similar way, we can estimate the optimal thresholds for 3-level quantization to be  $\boldsymbol{\beta}^*_3=[-0.341 \quad 0.528 ]$ as shown  in Figure  \ref{Entropy_threeLevel} in terms of equal level contours. Similarly, the optimum thresholds are $\boldsymbol{\beta}^*_4=[-0.367\quad 0.195 \quad 0.835 ]$ in the case of 4-level quantization and $\boldsymbol{\beta}^*_6=[-1.08\quad -0.572 \quad -0.060 \quad 0.4513 \quad 0.963 ]$ for 6-level quantization.

\subsection{Maximum J-divergence Method}

\begin{figure}
\centering
\includegraphics[width=12cm]{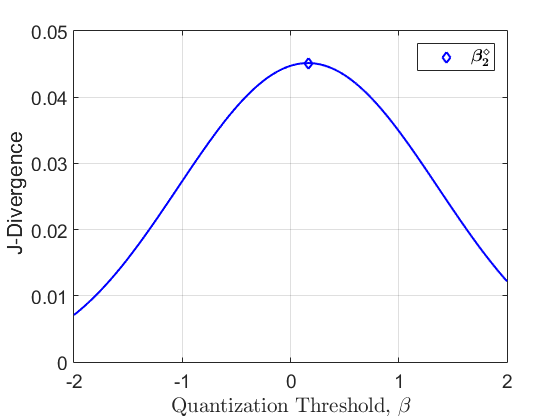}
\caption{ The J-divergence for binary quantization.}
  \label{Binary JD}
\end{figure}

\begin{figure}
\centering
\includegraphics[width=12cm]{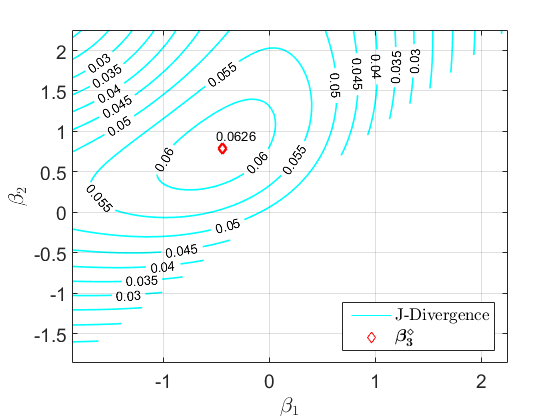}
\caption{The J-divergence for three level quantization.}
  \label{JD_threeLevel}
\end{figure}

J-divergence can be written  in terms of the relative entropy for discrete probability distributions $P$ and $Q$ observed under the two hypotheses $H_0$ and $H_1$ respectively as follows:

 \begin{equation}
\label{J-divergence1}
{J}=D_{KL}(P||Q)+D_{KL}(Q||P),
\end{equation}
where the relative entropy between two probability mass function (pmf) $P(x)$ and $Q(x)$ is given as follows:
\begin{equation}
\label{KLD-1}
D_{KL}(P||Q) =\sum_{x\in\chi}P(x)\log_2\left(\frac{P(x)}{Q(x)}\right), 
\end{equation}%
 where $\chi$ denotes the alphabet of the discrete pdfs for $P$ and $Q$. 
 In our context, $J$-divergence  measures the distributional distance or dissimilarity between the distributions of the observations under two hypotheses $H_0, H_1$ and this
 can be used  to find the  local thresholds. The choice of local thresholds facilitates the design of local detectors which in turn determines the performance of the whole system.
An estimate of the expected value for the {J} -divergence can be obtained by averaging the contribution to the J-divergence over the distribution of sensor locations and noise realizations as performed for entropy of the observations in $(\ref{lrcm})$ and can be written as:

\begin{equation}
\label{J -divergence2}
\hat{J}= \hat{E} \left(\sum_{m=1}^M\left[ p_m^{H_1} \log_2\left(\frac{p_m^{H_1}}{p_m^{H_0}}\right)- p_m^{H_0} \log_2\left(\frac{p_m^{H_1}}{p_m^{H_0}}\right)\right]\right).
\end{equation}
It is obvious  that $\hat{J}$ as specified by (\ref{J -divergence2}) is a function of the probability masses corresponding to the partitions of the pdf. For $M$-ary quantization, $\boldsymbol{\beta}^{{\diamond}}_M=[\beta^{\diamond}_1, \beta^{\diamond}_2, . . ., \beta^{\diamond}_{M-1}]$ denotes the J-divergence optimized vector of quantization thresholds which can be given as
\begin{equation}
\label{maximum JD}
\boldsymbol{{\beta}}^{\diamond}_M={\arg \max_{\boldsymbol{\beta}_M}} \quad {\hat{J}}.
\end{equation}
Optimal quantization thresholds correspond to the maximum of ${\hat{J}}$ which is found to be $\beta^{\diamond}_2=0.17$ for binary quantization as shown  in Figure  \ref{Binary JD}. In a similar  way, we can estimate the optimal thresholds for 3-level quantization to be $\boldsymbol{\beta}^{\diamond}_3=[-0.444 \quad 0.784 ]$ as shown  in Figure  \ref{JD_threeLevel} . Similarly, the optimal thresholds are $\boldsymbol{\beta}^{\diamond}_4=[-0.725\quad -0.699 \quad 0.6559 ]$ in the case of 4-level quantization and $\boldsymbol{\beta}^{\diamond}_6=[-6.19\quad -0.572  \quad -0.0603 \quad 0.9628 \quad 6.59 ]$ for the 6-level quantizations.


\subsection{Relation of MAE and MJD Methods}
 We will show that MAE and MJD obey a  positively proportional relation. Let us first express $D_{KL}(P||Q)$ given in (\ref{KLD-1}) as follows:
 
 \begin{equation}
\label{KLD0}
D_{KL}(P||Q) =\underbrace{\sum_{x\in\chi}P(x)\log_2(\frac{1}{Q(x)})}_{{\makebox[0pt]{$R_1$ }}}+\underbrace{\sum_{x\in\chi}P(x)\log_2(P(x))}_{{\makebox[0pt]{$-F_{H_0}$ }}}\geq 0.
\end{equation}
The equality holds only when  $P=Q$. Similarly,
\begin{equation}
\label{KLD1}
D_{KL}(Q||P)=R_2-F_{H_1} \geq 0,
\end{equation}
where $R_2=\sum_{x\in\chi}Q(x)\log_2(\frac{1}{P(x)}).$ Substituting (\ref{KLD0}) and (\ref{KLD1}) into (\ref{J-divergence1}) 
 \begin{equation}
\label{J-divergence1}
{J}=R_1+R_2-(F_{H_0}+F_{H_1})\geq 0.
\end{equation}
Then, defining $D_{KL}(P||Q)=c_1 F_{H_0} $ and  $D_{KL}(Q||P)=c_2 F_{H_1}$, we can re-write the $J-$divergence  in (\ref{J-divergence1}) to show that there is a proportionality relation between the JD and the average entropy (AE):
 \begin{equation}
\label{JD00}
\begin{split}
{J}&=c_1 F_{H_0}+c_2 F_{H_1}\\
&=\text{min}\{c_1,c_2\} \underbrace{(F_{H_0}+F_{H_1})}_{{\makebox[0pt]{$2 F_{\text{av}}$ }}}+c_3
\end{split}
\end{equation}
with
\begin{equation}
\label{constant C}
    {c_3}=\left\{
                \begin{array}{ll}
                                 (c_1-c_2)F_{H_0} \enspace \textnormal{for} \enspace c_1\geq c_2, \\
                                 (c_2-c_1)F_{H_1} \enspace \textnormal{for} \enspace c_1\leq c_2
                \end{array}
              \right.
 \end{equation}
Obviously $c_i\geq 0$ for $i=1,2\enspace \text{and} \enspace3$.

\clearpage

\section{Experimental Study}\label{Experimental Study}

Monte Carlo simulations have been performed in order to evaluate the detection performance for the proposed method at SNR$=0$ dB for $K=25$ transmitting sensors and $L=A_{max}/A_{min}=10$. First we have performed simulations using the direct data transmissions (DDT) method, that is assuming the sensor outputs are available error-free at the FC for both MAE and MJD methods. Then, a Rayleigh fading channel is considered to show the channel effect on the performance on our proposed quantization method, MAE. 

\subsection{Binary Direct Data Transmission}
\begin{figure}
\centering
\includegraphics[width=12cm]{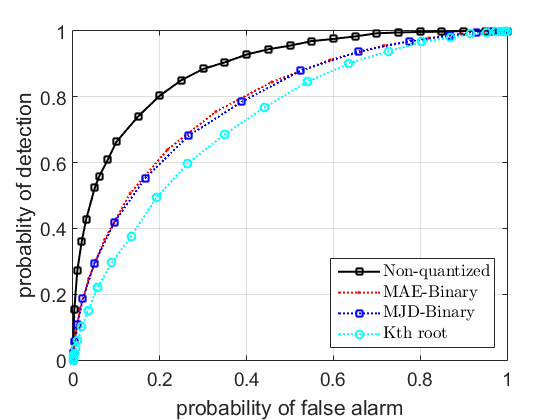}
\caption{Comparison between the ROC curves obtained using MAE, MJD and Kth root method for the binary data DDT transmissions.}
  \label{Binary_DDT_MAE_MJD_KthROOt}
\end{figure}

In Figure \ref{Binary_DDT_MAE_MJD_KthROOt}, the Receiver Operating Characteristics (ROC) curves are plotted for the cases of using the quantization intervals from MAE and MJD methods for the binary data transmission and the corresponding non-quantized data transmissions.
 The $K$th root quantization, which uses the $K$th root of the global probability of false alarm $p_\text{fa}=0.1$ to find pmf  $\tilde{p}_2^{H_0}=0.89$ at each sensor, is also provided for the comparison with the proposed method, MAE. $K$th root method corresponds to setting the false alarm threshold at the FC to a single "one" coming from any of $K$ sensors. In this figure, we observe a slightly better performance of MAE-based method compared to MJD-based one. Each of them perform much better compared to the trivial Kth root method which is supplied as an obvious lower bound. Additionally, we observe that they are clearly inferior to the non-quantized case which shows that there is quite a large space for gain in using higher levels of quantization.
\subsection{Performance of MAE and MJD with Multilevel Quantization}
\begin{figure}
\centering
\includegraphics[width=12cm]{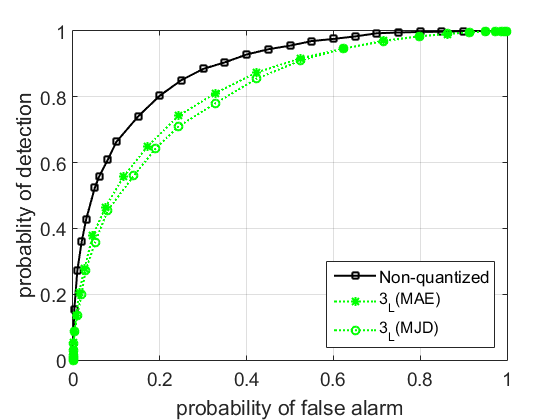}
\caption{Comparison between the ROC curves obtained using MAE, MJD method for the binary  DDT transmissions and Gaussian data transmissions.}
  \label{Three_DDT_G_MAE_MJD}
\end{figure}

The simulation performances for the three-level,  four-level and six-level quantizations by using the MAE and MJD methods are also obtained for DDT. 
 ROC curves obtained using MAE and MJD methods for three levels of quantization and non-quantized data are shown, in Figure \ref{Three_DDT_G_MAE_MJD}. This figure depicts that  at global false alarm probability $p_\text{fa}=0.2$, the probability of detection, $p_d$, attains the values $0.653, 0.684$ and $0.803$ for the cases of  three-level data transmissions with MJD, MAE and the non-quantized data transmission, respectively. Increasing the quantization level makes the MAE and MJD methods perform more closer to the performance without quantization which is depicted in Tables \ref{Comparison table2_3_4_6_G} and \ref{Gain table}. 

Table \ref{Comparison table2_3_4_6_G} shows the $p_{\text{d}}$ for $2$, $3$, $4$ and $6$ level MAE and MJD based  quantized  and non-quantized data transmissions for the values of $p_{\text{fa}}=0.1, 0.2, 0.3$ and $0.4$. At each quantization level MAE method performs better compared to MJD and the performance increases when the quantization level is increased. At $6$-level quantization $p_d$ obtained by MAE based method is only slightly inferior to the limiting case with no quantization, quantitatively, the difference in $p_d$ is $0.022$, $0.014$, $0.018$ and $0.002$ for $p_{fa}$ values of $0.1$, $0.2$, $0.3$ and $0.4$, respectively. Table \ref{Gain table} shows the achieved gain in $p_d$ by using the MAE method wrt MJD method and is given by $G=\left(p_d^{\text{MAE}_i}-p_d^{\text{MJD}_i}\right)$ with the resulting percentage gain $PG=(G/p_d^{\text{MAE}_i})\times 100\%$, where $i=2, 3, 4, 6$.
\newcolumntype{g}{>{\columncolor{Gray}}c}
\begin{table}
\setlength{\arrayrulewidth}{0.7mm}
\renewcommand{\arraystretch}{0.8}
    \begin{tabular}{|c|>{\columncolor[gray]{0.8}}c|>{\columncolor[gray]{0.8}}c|c|c|>{\columncolor[gray]{0.8}}c|>{\columncolor[gray]{0.8}}c|c|c|>{\columncolor[gray]{0.6}}c|}
    \hline
\diagbox{\shortstack{ $p_{\text{fa}}$}}
        {\shortstack{\vphantom{Ä}$p_{\text{d}}$}} & $\text{MJD}_2$ & $\text{MAE}_2$ &  $\text{MJD}_3$ & $\text{MAE}_3$ & $
\text{MJD}_4$ & $\text{MAE}_4$ & $\text{MJD}_6$ & $\text{MAE}_6$ & $\text{non-quantized}$\\ \hline
$0.1$ &0.425&0.432 &0.497&0.520&0.567	&0.592&0.629		&0.643&0.665\\ \hline
$0.2$ &0.590&0.610 &0.653&0.684&0.728	&0.760&0.772		&0.789&0.803\\ \hline
$0.3$ &0.710&0.720 &0.755&0.790&0.810	&0.845&0.850		&0.867&0.885\\ \hline
$0.4$ &0.787&0.805 &0.825&0.858&0.860	&0.895&0.903		&0.922&0.924 \\ \hline
  \end{tabular}\\
\caption{The relation between $p_d$ and $p_\text{fa}$ for the different level of quantization obtained with MAE and MJD methods.}
\label{Comparison table2_3_4_6_G}
\end{table}
\begin{table}
\setlength{\arrayrulewidth}{0.7mm}
\renewcommand{\arraystretch}{0.8}
\begin{tabular}{%
   |>{\Centering}m{1.5cm}
    |*{9}{>{\Centering}m{3em}|}}\hline
\diagbox{\shortstack{ $p_{\text{fa}}$}}
        {\shortstack{\vphantom{Ä}$p_{\text{d}}$}} &\multicolumn{2}{l|}{$2$-Level}&\multicolumn{2}{l|}{$3$-Level}&\multicolumn{2}{l|}{$4$-Level}&\multicolumn{2}{l|}{$6$-Level}\\
\cline{2-9}
 &G&PG&G&PG &G&PG &G&PG\\
\hline\hline
$0.1$ &0.01 &1.63  &0.023  &4.42  &0.025  &4.22  &0.014&2.17\\ \hline
$0.2$ &0.02&3.28&0.031&4.53&0.032&4.21&0.017 &2.15\\ \hline
$0.3$ &0.01&1.39&0.035&4.43&0.035&4.14&0.017&1.96\\ \hline
$0.4$ &0.018&2.2&0.0330&3.85&0.035&3.914&0.019&2.06\\ \hline
  \end{tabular}\\
\caption{Achieved gain in $p_d$ by using the MAE method in quantization instead of MJD.}
\label{Gain table}
\end{table}

It is obviously seen from the previous tables that MAE outperforms MJD for $M\geq2$ levels. The achieved gain of MAE wrt MJD is in average $0.0138$ with a corresponding percentage gain of $2.13\%$ for the binary data transmissions, whereas the average gains are $=0.0305$, $0.0318$ and $0.0168$ with corresponding average percentage gains as $4.31\%$, $4.12 \%$ and $2.09 \%$ for 3-level, 4-level and 6-level data transmissions, respectively. In the same manner, the average difference in $p_d$, for $p_\text{fa}=0.1, 0.2, 0.3$ and $0.4$, between the $6$-level data transmissions achieved by MAE and non-quantized data transmissions equals to $0.014$ with $1.8\%$ it is $0.03$ with $3.9\%$ between MJD and non-quantized data transmissions, which result shows that $6$-level data transmission by using MAE is very close to the non-quantized data transmission and gives better performance than MJD method.

\subsection{Multiple Level Data Transmission over Rayleigh Fading Channel}

\begin{figure}
\centering
\includegraphics[width=12cm]{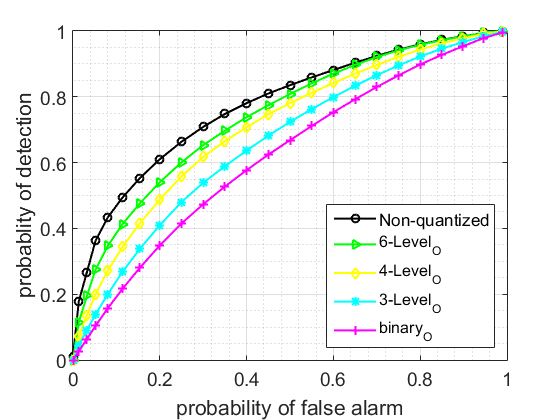}
\caption{ROC curves by using MAE quantization over fading channel.}
  \label{Rayleigh_Optimal_fading_channelV5}
\end{figure}

\begin{figure}
\centering
\includegraphics[width=12cm]{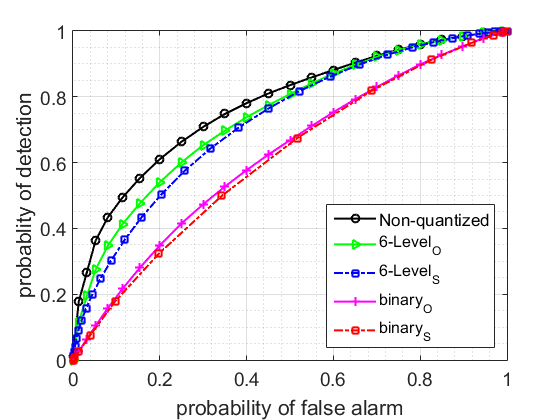}
\caption{ A comparison between the ROC of the optimal and sub optimal fusion rule for binary and six level data transmissions and its corresponding non-quantized data transmissions.}
  \label{Rayleigh_SUB_Optimal_fading_channelV5}
\end{figure}

Figure \ref{Rayleigh_Optimal_fading_channelV5} shows the ROC curves for $2$, $3$, $4$ and $6$ level MAE based quantized and non-quantized data transmissions by using $M$-FSK modulation scheme with non-coherent demodulation over Rayleigh fading channels and AWGN by using the optimal fusion rule in (\ref{fusion rule v3}), the threshold, $\eta$, estimated by running a Monte Carlo simulation under no event case and finds thresholds corresponding to each $p_\text{fa}$. In this figure we can see that the obtained, $p_d$ for $6$-level quantization underperforms the limiting case of no quantization by $0.09$ at $p_\text{fa}=0.1$. This gain diminishes at $p_\text{fa}=0.7$. When we compare the performances of different quantization levels, the achieved gain in $p_d$ by transmitting $6$-level quantization instead of  $2$-level quantization is $0.21$ for $p_\text{fa}=0.1$ and this gain diminishes at $p_\text{fa}=0.99$. Also, the sub-optimal fusion rule in (\ref{finalDecsion})  have been used to find the ROCs for the different type of data transmissions. Figure \ref {Rayleigh_SUB_Optimal_fading_channelV5} shows a comparison between the optimal and sub-optimal fusion rule for $2$ and $6$ levels data transmissions and compare them with the non-quantized data transmissions. The dashed line in ROCs for the sub-optimal fusion rule correspond to randomization in the tests \cite{poor}. From this figure we can see the achieved gain by using the optimum fusion rule wrt the sub-optimal rule is $0.3$ and $0.6$ at $p_\text{fa}=0.1$ for the $2$-level and $6$- level data transmissions, respectively.

\clearpage
\clearpage

\section{Conclusion}\label{section:Conclusion}
In this study we have proposed quantizing the sensor outputs by maximizing their average information in the cases of presense and non-presense of an event in decentralized detection. 

The general approach in quantization for decision processes is based on distance measures such as J-divergence and Bhattacharyya distance. This fact may have prevented a popular information based quantization criterion for decision processes maximizing the information under both (all) hypotheses rather than the information in the difference of the distributions. Since among the distance measure based quantization approaches, J-divergence is an information theoretic quality, we adopted J-divergence for comparisons of  the proposed method.

One reason of suggesting another method like MAE instead of MJD might be the non-symmetric nature of the considered problem and that the advantage of Ali-Silvey type criteria \cite{poor} which MJD is a member of, is only valid for the symmetric performance measure probability of error. Although maximizing the transferred information under each hypothesis as proposed by the MAE method is a  conceptually different approach, we showed that average entropy and J-divergence are actually positively proportional quantities. This means that one might expect comparable performances using either of them for determining the quantization levels which was indeed the observation in the simulation results.

In order to isolate  the effects of how the sensor outputs are quantized on the system performance we performed extensive simulation studies for the case that the sensor outputs are available error-free at the FC which we called as DDT. The performances of considered information-based methods, namely MAE and MJD, gradually as the quantization level was increased from binary to six-levels and it approached the performance of non-quantized data transmission. Additionally, the proposed method, MAE, performed significantly better compared to MJD for any level of quantization. Also, the effects of Rayleigh fading channel from the sensors to the FC have been investigated using the optimal and a suboptimal fusion rule for MAE. Due to the power efficiency and small degradation in non-coherent communication MFSK was adopted as the modulation scheme for the sensor to FC communication. Using the wireless channel model similar results were obtained as in DDT. Results with 6-level quantization were comparable to non-quantized data transmission. 

This work showed that MAE is a valid and promising method in quantization for detection problems. a possible future work will be applying MAE method of quantization in M-ary detection problems.


\end{document}